\documentclass[copyright,creativecommons]{eptcs}
\providecommand{\event}{GASCom 2024} 

\usepackage{iftex}

\ifpdf
  \usepackage{underscore}         
  \usepackage[T1]{fontenc}        
\else
  \usepackage{breakurl}           
\fi

\usepackage{amsmath,amssymb,amsthm}
\usepackage{mathtools}
\usepackage{url}
\usepackage{hyperref}
\usepackage{shuffle}
\usepackage{wasysym}

\usepackage[linesnumbered,boxed,vlined]{algorithm2e}	
\SetAlCapSkip{0.5em}

\usepackage{graphicx}
\theoremstyle{definition}

\theoremstyle{plain}
\newtheorem{theorem}{Theorem}[section]
\newtheorem{prop}[theorem]{Proposition}

\newtheorem*{conj}{Conjecture}

\DeclareMathOperator{\bub}{\mathbf{B}}
\DeclareMathOperator{\que}{\mathbf{Q}}

\DeclareMathOperator{\psb}{\mathbf{PSB}}

\DeclareMathOperator{\stack}{\mathbf{S}}

\newcommand{\Av}{\mathrm{Av}}

\title{Pop Stacks with a Bypass}
\author{Lapo Cioni\thanks{Partially supported by INdAM -- GNCS group.}
\institute{Dipartimento di Informatica\\ University of Pisa, Pisa, Italy}
\email{lapo.cioni@di.unipi.it}
\and
Luca Ferrari\thanks{Partially supported by INdAM -- GNCS project CUP E53C23001670001.}
\institute{Dipartimento di Matematica e Informatica ``U. Dini”\\ University of Firenze, Firenze, Italy}
\email{luca.ferrari@unifi.it}
\and
Rebecca Smith
\institute{Department of Mathematics\\ SUNY Brockport, Brockport, New York}
\email{rnsmith@brockport.edu}
}
\def\titlerunning{Pop stacks with a bypass}
\def\authorrunning{L. Cioni, L. Ferrari \& R. Smith}
\begin{document}
\maketitle

\begin{abstract}
	We consider sorting procedures for permutations making use of pop stacks with a bypass operation, and explore the combinatorial properties of the associated algorithms.
\end{abstract}

\section{Introduction}

Sorting techniques for permutations constitute a flourishing research area in contemporary combinatorics. Starting from the stack-sorting procedure introduced by Knuth \cite{K}, over 150 articles (many of which are quite recent) have explored the subject. For example, machines making use of various types of containers have been studied, as well as networks of the corresponding devices. Typical containers that are considered in this context include stacks, queues, and deques, as well as their corresponding ``pop" versions. In particular, a \emph{pop stack} is a stack whose push and pop operations are similar to the usual ones for stacks, except that a pop operation extracts \emph{all} the elements from the stack, rather than just the element on the top. The sorting power of a pop stack was originally studied by Avis and Newborn \cite{AN}, who also considered pop stacks in parallel. Concerning pop stacks in series, we mention that the right-greedy version of pop stacks in series were introduced by Pudwell and Smith \cite{PS} and successively studied by Claesson and Gu\dh{}mundsson \cite{CG}.

In this work, we consider a new variant of a pop stack, in which we also allow a bypass operation. This gives the possibility of sending an element of the input permutation directly into the output, without necessarily pushing it into the pop stack first. This increases the sorting power of a pop stack, and inspires interesting combinatorial questions on its properties. We provide a detailed analysis of this sorting device, specifically our results are the following:

\begin{itemize}
	\item We characterize the set of sortable permutations in terms of two forbidden patterns; the enumeration of the resulting class of pattern avoiding permutations was already known (odd-indexed Fibonacci numbers, sequence A001519 in \cite{Sl}). However, we give an independent proof of this enumerative result by describing a bijective link with a restricted class of Motzkin paths;
	\item We describe an algorithm to compute the preimage of a given permutation, and use it to characterize and enumerate permutations having exactly 0, 1, and 2 preimages;
	\item We provide a complete description of the preimages of principal classes of permutations, determine in which cases the preimages are classes, and in such cases we compute the basis of the resulting classes;
	\item We characterize the set of sortable permutations for the compositions of our sorting algorithm with other classical sorting algorithms (the characterization being given in terms of forbidden patterns);
	\item We consider the device consisting of two pop stacks in parallel with a bypass, and we determine the basis of the associated class of sortable permutations.
\end{itemize}

\section{Sorting using a pop stack with a bypass: characterization and enumeration of sortable permutations}

A \emph{pop stack} is a container in which elements can be stacked on top of each other, on which two operations are defined: the \texttt{PUSH} operation, which inserts an element in  the stack, and the \texttt{POP} operation, which extract \emph{all the elements} from the stack. The difference between a pop stack and a (classical) stack lies therefore in the way the elements are removed from the stack.

When using a stack or a pop stack to sort a permutation, the elements of the permutation are processed from left to right, and either the current element of the permutation is pushed into the stack at the top, or the topmost element (or all the content from top to bottom in the case of a pop stack) of the stack is popped into the output, where the topmost element is placed to the right of any other elements already in the output. We now introduce a new kind of pop stack, by allowing an additional \texttt{BYPASS} operation, which takes the current element of the permutation and places it directly into the output, in the next available position. More formally, given a permutation $\pi=\pi_1 \pi_2 \dots \pi_n$, a pop stack with bypass has the following allowed operations:
\begin{itemize}
	\item[]\texttt{PUSH:} Insert the current element of the input into the pop stack, on top of all the other elements (if there are any);
	\item[]\texttt{POP}: Remove all the elements in the pop stack, from top to bottom, sending them into the output;
	\item[]\texttt{BYPASS}: Output the current element of the input.    
\end{itemize}

Our goal is to use a pop stack with a bypass to sort permutations. However, not every permutation can be sorted. Algorithm \ref{Popstackbypass} (see below), called \textsf{PSB} (an acronym for ``PopStacksort with Bypass"), is an optimal sorting algorithm, by which we can sort all sortable permutations.

\begin{algorithm}\label{Popstackbypass}
	$S:=\emptyset$\;
	$i:=1$\;
	\While{$i\leq n$}
	{
		\If{$S=\emptyset$ or $\pi_i =\textnormal{\texttt{TOP}}(S)-1$}
		{
			\texttt{PUSH}\;
		}
		\ElseIf{$\pi_i <\textnormal{\texttt{TOP}}(S)-1$}
		{
			\texttt{BYPASS}\;
		}
		\Else
		{
			\texttt{POP}\;
			\texttt{PUSH}\;
		}
		$i:=i+1$\;		
	}
	\texttt{POP}\;
	\caption{\textsf{PSB} ($S$ is the pop stack; \texttt{TOP}(S) is the current top element of the pop stack; operations \texttt{PUSH}, \texttt{POP}, \texttt{BYPASS} are ``Insert into the pop stack", ``Pour the whole content of the pop stack into the output", ``Bypass the pop stack", respectively;  $\pi=\pi_1 \cdots \pi_n$ is the input permutation.)}\label{queuesort}
\end{algorithm}

Specifically, the algorithm \textsf{PSB} maintains elements in the pop stack only when they are consecutive in value (and increasing from top to bottom), which ensures that a \texttt{POP} operation is not directly responsible of the possible failing of the sorting procedure. Using the algorithm \textsf{PSB}, we can characterize the class of sortable permutations. Denote by $\psb$ the map associated with the algorithm \textsf{PSB}. Moreover, given a set of permutations $T$, we indicate with $\Av_n (T)$ the set of permutations of size $n$ avoiding all patterns of the set $T$, and we write $id_n$ for the identity permutation of size $n$. As usual, the set of all permutations of size $n$ is denoted $S_n$.

\begin{prop}
	Given $\pi \in S_n$, we have that $\psb (\pi )=id_n$ if and only if $\pi \in \Av_n (231,4213)$.   
\end{prop}

The class of permutations sortable using a popstack with a bypass is a superclass of those sortable using a classical popstack, which is $\Av_n (231,312)$, as shown in \cite{AN}. The sequence $(|\Av_n (231,4213)|)_n$ is the sequence of odd-indexed Fibonacci numbers, i.e. sequence A001519 in \cite{Sl}, as shown in \cite{At}. It is possible to give a bijective proof of this enumerative result by providing a link with a class of restricted Motzkin paths, whose enumeration can be easily carried out by standard techniques. The next proposition describes which kinds of restrictions on Motzkin paths need to be considered.  

\begin{prop}
	The set of permutations of size $n$ sortable using a pop stack with a bypass is in bijection with the set of Motzkin paths whose total number of up and horizontal steps is $n$, which have no peaks, and are such that every maximal sequence of down steps reaches the bottom level.   
\end{prop}

\section{Preimages of a permutation}

In this section, we investigate the set of permutations $\psb^{-1}(\sigma)$ of all permutations whose output under \textsf{PSB} is $\sigma$. We remark that the investigation of preimages of permutations under various sorting operators was initiated for Stacksort in \cite{B-M}, and is now a fertile area of research.

We introduce a recursive algorithm that generates all preimages of a given permutation $\sigma$.  
Notice that our algorithm is actually defined for any sequence of distinct integers (not only permutations).  This fact allows us to recursively execute it on subsequences of elements of a permutation. Recall that a \emph{left-to-right maximum} of a permutation is an element which is larger than all elements to its left. Moreover, we say that two entries of $\pi$ are \emph{consecutive} when their \emph{values} are consecutive integers, whereas we say that they are \emph{adjacent} when their \emph{positions} are consecutive integers.

\medskip

Suppose that $\sigma =\sigma_1 \sigma_2 \cdots \sigma_n =\alpha \mu_k$, where $\mu_k$ is the maximum suffix of consecutive left-to-right maxima of $\sigma$ (and $\alpha$ is the remaining prefix). For each entry $m$ in $\mu_k$, construct permutations as follows. First, remove the suffix of left-to-right maxima starting with $m$. Then reinsert the removed elements into the (remaining prefix of the) permutation in all possible ways, according to the following rules:
\begin{itemize}
	\item The removed elements are reinserted in decreasing order;
	\item The maximum (i.e. $n$) is inserted to the immediate right of one of the remaining left-to-right maxima of $\sigma$ (i.e. a left-to-right maximum to the left of $m$ in $\sigma$)
	or at the beginning of $\sigma$;
	\item The minimum (i.e. $m$) is inserted somewhere to the right of $m-1$. 
\end{itemize}

At this point, consider the prefix of all elements strictly before $n$ and (recursively) compute all its possible preimages. 

\medskip

Using this algorithm we are able to characterize and enumerate permutations having a small number of preimages.

Given $n\ge 0$, let $C_n^{(k)}=\{\sigma \in S_n\mid |\psb^{-1}(\sigma )| = k\}$, and let $c_n^{(k)}=|C_n^{(k)}|$.

\begin{prop}
	\[
	C_n^{(0)}=\{\sigma =\sigma_1 \cdots \sigma_n \mid \sigma_n \neq n\}
	\]
	and 
	\[
	c_n^{(0)}=(n-1) (n-1)!.
	\]
\end{prop}

In the subsequent cases, for a given permutation $\pi$, we let $LTR(\pi)$ denote the set of the left-to-right maxima of $\pi$.

\begin{prop}
	For $n\geq 3$, the set $C_n^{(1)}$ consists of all permutations of size $n$ ending with $n$ whose left-to-right maxima are consecutive and nonadjacent. More formally,  
	\begin{gather*}\label{preim1}
	C_n^{(1)}=\{\pi=\pi_1\cdots\pi_n\mid \pi_n=n \text{ , there exists $k\ge 0$ such that } LTR(\pi)=\{n-k,\dots,n\}\\
	\text{ and for every $\pi_i$, $\pi_j \in LTR(\pi)$, $\pi_i\neq\pi_j$, we have } |j-i|>1 \} .
	\end{gather*}
	
	Moreover
	\[
	c_n^{(1)} = \sum_{k=2}^{\lceil\frac{n}{2}\rceil} (n-k)! \binom{n-k-1}{k-2}.
	\]
\end{prop}

The first terms of the sequence $c_n^{(1)}$ (starting from $n=1$) are $1,0,1,2,8,36,198,\ldots$, and do not appear in the OEIS~\cite{Sl}.

\begin{prop}
	For $n\geq 4$, the set $C_n^{(2)}$ consists of all permutations of size $n$ ending with $n$ whose left-to-right maxima are consecutive and nonadjacent	\emph{except for the first one}, which \emph{is required} to be nonconsecutive with the second one, and \emph{can possibly} be adjacent to the second one. More formally,	
	\begin{gather*}
	C_n^{(2)}=\{\pi=\pi_1\cdots\pi_n\mid \pi_n=n \text{, there exists $k\ge 0$ such that } LTR(\pi)=\{\pi_1, n-k,\dots,n\} ,\\
	\text{ with } \pi_1\neq n-k-1, \text{ and for every $\pi_i$, $\pi_j \in LTR(\pi)$, $\pi_i\neq\pi_j$, $i,j\neq 1$, we have } |j-i|>1 \}.
	\end{gather*}
	
	Moreover
	\begin{equation}
	c_n^{(2)} = \sum_{k=3}^{n} \sum_{j=1}^{n-k} \frac{n-k-j+1}{j}(n-k)!\binom{n-j-k}{k-3}.
	\end{equation}
\end{prop}

\section{Preimages of classes}

Next we consider the preimages of principal classes under $\psb$, and we completely determine the permutations $\rho$ for which $\psb^{-1}(\Av (\rho ))$ is a class. Moreover, in all cases we explicitly describe the basis of the resulting class. Our results are also useful in the next section, where we will compose \textsf{PSB} with other sorting algorithms.  

We recall that a \emph{permutation class} (or simply \emph{class}) is a set $C$ of permutations for which every pattern contained in a permutation in $C$ is also in $C$. Every permutation class can be defined by the minimal permutations which do not lie inside it, its \emph{basis}. A \emph{principal permutation class} is a class whose basis consists of a single permutation.

In order to properly describe our results, we will use the notion of shuffle.
Let $\rho$, $\sigma$ be two sequences of distinct integers. Then we say that a sequence $\tau$ is a \emph{shuffle} of $\rho$ and $\sigma$ if $\tau$ contains both $\rho$ and $\sigma$ as subsequences, and contains no elements other than those of $\rho$ and $\sigma$.
We denote the set of all possible shuffles of $\rho$ and $\sigma$ with $\rho\shuffle\sigma$.

\begin{prop}
	Let $\rho$ be a permutation.
	\begin{itemize}
		\item If $\rho =\emptyset ,1$ or $12$, then we have that $\psb^{-1}(\Av (\rho ))= \Av (\emptyset ),\Av (1)$, or $\Av (12,21)$, respectively.
		\item If $\rho=n\alpha$, for some $\alpha \in S_{n-1}$, $\alpha\neq\emptyset$, then $\psb^{-1}(\Av(\rho))=\Av(B)$, where \\ $B=\{ n(n+1)\alpha \} \cup \{ (n+2)n\tau \, |\, \tau \in (n+1)\shuffle \alpha , \tau \neq (n+1)\alpha \}$.
		\item If $\rho=(n-1)\alpha n$, for some $\alpha \in S_{n-2}$, $\alpha\neq\emptyset$, then $\psb^{-1}(\Av(\rho))=\Av(B)$, where \\$B=\{ (n-1)n\alpha \} \cup \{ (n+1)(n-1)\tau \, |\, \tau \in n\shuffle \alpha , \tau\neq n\alpha \}$.
		\item In all the remaining cases, $\psb^{-1}(\Av(\rho))$ is not a permutation class.
	\end{itemize} 
\end{prop}

\section{Composition with other sorting algorithms}

If we execute a sorting algorithm \texttt{X} on a permutation and then execute another sorting algorithm \texttt{Y} on the output permutation, we get a new sorting algorithm, whose associated map is the \emph{composition} of the maps associated with \texttt{X} and \texttt{Y}.   

In this section, we investigate permutations which are sortable by a composition of \textsf{PSB} with another sorting algorithm. Specifically, we will consider the algorithms \texttt{Queuesort}, \texttt{Stacksort}, and \texttt{Bubblesort}, whose associated maps will be denoted $\que$, $\stack$, and $\bub$, respectively (recall that the algorithm \texttt{Queuesort} uses a queue with a bypass). Our results are summarized in the next proposition.

\begin{prop}
	Let $\pi \in S_n$.
	\begin{itemize}
		\item $\stack\circ\psb(\pi)=id_n$ if and only if $\pi\in\Av(2341, 25314, 52314,$ $45231,$ $42531,3\bar{5}241)$.
		\item $\que\circ\psb(\pi) = id_n$ if and only if $\pi\in\Av(3421, 53241, 53214)$.
		\item $\bub\circ\psb(\pi)=id_n$ if and only if $\pi\in\Av(2341, 3421, 3241,$ $25314, 52314, 53214)$.
		\item $\psb \circ \que (\pi ) = id_n$ if and only if $\pi \in \Av (4231, 2431, 54213)$.
		\item $\psb\circ\bub(\pi) = id_n$ if and only if $\pi\in\Av(2341,2431,3241,$ $4231,45213,54213)$.
	\end{itemize}
\end{prop}

\section{Pop stacks in parallel with a bypass}

We now consider a sorting machine that consists of two pop stacks in parallel, where an entry can bypass the pop stacks and instead go directly to the output.  Pop stacks in parallel without a bypass were introduced by Avis and Newborn~\cite{AN}, as already recalled in the Introduction, and later studies include that of Atkinson and Sack~\cite{AS} and Smith and Vatter~\cite{SV}.

We now informally describe an algorithm to sort a permutation using the above described device. Since we are not interested in composing it with other sorting algorithm, our procedure will provide an output only when the sorting succeeds (otherwise the algorithm just fails). 

Let $\pi =\pi_i\pi_2\cdots\pi_n$ be a permutation in the input at the start. Let $S_1, S_2$ be the two pop stacks. Suppose that $\pi_i$ is the next entry in the input. 
\begin{itemize}
	\item If $\pi_i$ is the next entry needed in the output to obtain the identity, push $\pi_i$ to the output (i.e. $\pi_i$ bypasses the pop stacks).
	\item Else, if the top entry of $S_j$  (for either $j=1$ or $j=2$) is the next entry needed in the output to obtain the identity, pop the contents of $S_j$.
	\item Else, if $\pi_k$ is the top entry in the pop stack $S_j$ (for either $j=1$ or $j=2$) where $\pi_i=\pi_{k_i}-1$, push $\pi_i$ into $S_j$.
	\item Else, if one of the pop stacks is empty, push $\pi_i$ into an empty pop stack.
	\item Else, the sorting algorithm fails.
\end{itemize}

\medskip

It is possible to prove that the above algorithm
is optimal, as usual in the sense that it sorts all permutations sortable by the machine consisting of two pop stacks in parallel with a bypass option. Using the above procedure, we are able to characterize sortable permutations in terms of forbidden patterns.

\begin{prop}
	The class of permutations sortable by two pop stacks in parallel with a bypass is 
	$$\Av (2341,25314,42513,42531,45213,45231,52314,642135,642153).$$
\end{prop} 

As correctly pointed out by one referee (whom we warmly thank), the inverse of the above class (i.e. the class whose basis is the set of inverses of the above permutations) has a regular insertion encoding \cite{V}, thus it is possible to automatically deduce its rational generating function \cite{BEMNPU}, which is
\[
\frac{(1-x)(1-2x)(1-4x)}{1-8x+20x^2 -18x^3 +3x^4}.
\]   

The sequence corresponding to the sortable permutations of size $n$ has first terms $1,2,6,23,97,$ $418,1800,7717,32969,140558,\ldots$ and does not appear in \cite{Sl} at the time of this writing.  There is a potential match for the simple sortable permutations that we boldly put here as a conjecture:

\begin{conj}
	Let $a_n$ be the number of simple permutations of size $n$ which are sortable by a machine consisting of two pop stacks in parallel where entries are allowed to bypass the pop stacks. Then \\ $a_0 =a_1 =1$, $a_2 =2$, $a_n =F_{2n-5}-1$ if $n\geq 3$ is odd, and $a_n =F_{2n-5}$ if $n > 3$ is even (where $F_n$ is the $n$-th Fibonacci number).
\end{conj}

\bibliographystyle{eptcs}
\bibliography{generic}
\end{document}